# Q & A EXPERIMENT TO SEARCH FOR VACUUM DICHROISM, PSEUDOSCALAR-PHOTON INTERACTION AND MILLICHARGED FERMIONS

SHENG-JUI CHEN, HSIEN-HAO MEI, WEI-TOU NI

*Center for Gravitation and Cosmology, Department of Physics,*
*National Tsing Hua University, Hsinchu, Taiwan, 30013 Republic of China*
*wtni@phys.nthu.edu.tw*

A number of experiments are underway to detect vacuum birefringence and dichroism --- PVLAS, Q & A, and BMV. Recently, PVLAS experiment has observed optical rotation in vacuum by a magnetic field (vacuum dichroism). Theoretical interpretations of this result include a possible pseudoscalar-photon interaction and the existence of millicharged fermions. Here, we report the progress and first results of Q & A (QED [quantum electrodynamics] and Axion) experiment proposed and started in 1994. A 3.5-m high-finesse (around 30,000) Fabry-Perot prototype detector extendable to 7-m has been built and tested. We use X-pendulums and automatic control schemes developed by the gravitational-wave detection community for mirror suspension and cavity control. To polarize the vacuum, we use a 2.3-T dipole permanent magnet, with 27-mm-diameter clear borehole and 0.6-m field length,. In the experiment, the magnet is rotated at 5-10 rev/s to generate time-dependent polarization signal with twice the rotation frequency. Our ellipsometer/polarization-rotation-detection-system is formed by a pair of Glan-Taylor type polarizing prisms with extinction ratio lower than $10^{-8}$ together with a polarization modulating Faraday Cell with/without a quarter wave plate. We made an independent calibration of our apparatus by performing a measurement of gaseous Cotton-Mouton effect of nitrogen. At present, the sensitivity (and noise floor) for dichroism detection is about $1 \times 10^{-6}$ rad Hz$^{-1/2}$ at 10-20 Hz. With this sensitivity, it would be possible to check the polarization rotation effect obtained by PVLAS. Our first results give $(-0.2 \pm 2.8) \times 10^{-13}$ rad/pass, at 2.3 T with 18,700 passes through a 0.6 m long magnet for vacuum dichroism measurement. We present a brief discussion of our experimental limit on pseudo-scalar-photon interaction and millicharged fermions.

## 1. Introduction

Experiments to detect vacuum dichroism and vacuum birefringence have been of current focus since the recent positive result of PVLAS (Polarizzazione del Vuoto con LASer).[1] BFRT (Brookhaven-Fermilab-Rochester-Trieste) experiment[2] was completed in 1992, using superconducting magnet and multipass cavity, and put a limit of $(0.45\pm0.38) \times 10^{-14}$ rad T$^{-2}$ m$^{-2}$ (λ: 514 nm) on possible optical rotation of beam polarization in vacuum in a magnetic field. After BFRT experiment, 3 experiments were put forward for measuring the vacuum birefringence in 1994: the PVLAS experiment,[3] the Q & A (QED [quantum electrodynamics] & Axion) experiment,[4] and the Fermilab P-877 experiment;[5] these experiment were reported in the "Frontier Test of QED and Physics of Vacuum" Meeting in 1998. Fermilab P-877 experiment was terminated in 2000. However, BMV (Biréfringence Magnétique du Vide) experiment[6] started in 2000. The basic characteristics of the PVLAS experiment, the Q & A experiment, and the BMV experiment for measurement of the vacuum birefringence and dichroism are compiled in Table 1.[7] All 3 experiments use high-finesse Fabry-Perot Interferometer (FPI) cavity to enhance the effect to be measured. Recently, PVLAS experiment[1] reported a positive measurement of polarization rotation in vacuum by a magnetic field (vacuum dichroism) and suggested a possible interpretation of this result to the existence of a pseudoscalar interacting with photon.

Table 1. A comparison of the vacuum birefringence and dichroism experiments – PVLAS, Q & A and BMV experiments.

| Experiment | PVLAS | Q & A | BMV |
| --- | --- | --- | --- |
| Status | achieved | Achieved(2002)/Achieved(2006)/*goal* | Achieved/2007 goal/*goal* |
| Wavelength (nm) | 1064 | 1064 / 1064 / *532* | 1064 |
| Type of dipole magnet | Rotating superconducting | Switching / Rotating permanent /*Rotating permanent* | Pulsed |
| Magnetic field B (T) | 5.5 | 1.2 / 2.3 / *2.5* | 6 / 10 / *25* |
| Length of magnetic field (m) | 1 | 0.21 / 0.6 / *5* | 0.3 / 0.3 / *1.5* |
| Finesse of Fabry-Perot Interferometer | 70,000 | 11,620 / 31,000 / *100,000* | 50,000/ 200,000 / *10$^6$* |
| Magnet field modu-lation frequency (Hz) | 0.3 | 0.05 / 7.06 / *10* | 37 |
| Sensitivity (rad·Hz$^{-0.5}$) | ~10$^{-7}$ | 5·10$^{-6}$ / 5·10$^{-7}$ / *10$^{-8}$* | -- / < 10$^{-8}$ / *10$^{-9}$* |

(Pseudo)scalar-photon interaction was first proposed in 1973-4 under the study of the relationship of Galileo equivalence principle and Einstein equivalence principle for electromagnetic systems.[8-10] The Lagrangian density of this theory in gravitational field is given by

$$L_I = -(1/16\pi)g^{ik}g^{jl}F_{ij}F_{kl} - (1/16\pi)\varphi F_{ij}F_{kl}e^{ijkl} - A_k j^k (-g)^{(1/2)} - \Sigma_I m_I (ds_I)/(dt)\delta(\mathbf{x}-\mathbf{x}_I), \quad (1)$$

where $g_{ij}$ is the metric, $g$ its determinant, $j^k$ the charge current, $F_{ij} \equiv A_{j,i} - A_{i,j}$ the electromagnetic field, $\varphi$ the (pseudo)scalar field, and $e^{ijkl}$ the completely antisymmetric symbol with $e^{0123} = 1$.[8-10] The second term of (1)

$$L_I = -(1/16\pi)\varphi F_{ij}F_{kl}e^{ijkl} \quad (2)$$

gives the (pseudo)scalar interaction with photons. The effect of $\varphi$ in this theory on photon propagation is to change the phases of two different circular polarizations and gives polarization rotation for linearly polarized light.[8, 11, 12] The current limit on the variation of $\varphi$ over observed cosmological distance is about 0.1 from cosmic microwave background observations.[13-15]

In the study of strong CP problem, Weinberg[16] and Wilczek[17] proposed axion based on the work of Peccei and Quinn.[18] Their MeV axion model did not agree with accelerator experimental data. Soon the 'invisible' axion models with small mass were proposed.[19-21] In these axion models the photon-axion interaction Lagrangian is of the form (2). In terms of Feynman diagram, the interaction (2) gives a 2-photon-pseudo-scalar vertex. With this interaction, vacuum becomes birefringent and dichroic.[22-24]

Dichroic materials have the property that their absorption constant varies with polarization. When polarized light goes through dichroic material, its polarization is rotated due to difference in absorption in 2 principal directions of the material for the 2 polarization components. For axion models, the polarization rotation ε of the photon beam for light entering the magnetic-field region polarized at an angle of $\theta$ to the magnetic field is

$$\varepsilon = (B^2\omega^2 M^{-2} m_\varphi^{-4}) \sin^2(m_\varphi^2 L/4\omega) \sin(2\theta) \approx B^2 L^2/(16M^2) \sin(2\theta), \quad (3)$$

where $m_\varphi$ is mass of the axion, $M$ is the axion-photon interaction energy scale, $\omega$ photon circular frequency and $L$ the magnetic-region length. The approximation is valid in the limit

$$m_\varphi^2 L/4\omega \ll 1, \quad (4)$$

i.e., for a given length $L$ of the magnetic field region,

$$m_\varphi^2 < 2\pi\omega/L, \quad (5)$$

to preserve the the relative phase of the axion and photon fields. Since axions do not reflect at the mirrors, for multi-passes, the rotation effect increases by number $N$ of passes. Therefore for the case condition (5) is satisfied, the polarization rotation effect is proportional to $NB^2L^2$.

In Table 2, we compile the experimental results of three experiments on the vacuum dichroism – BFRT, PVLAS and Q & A.

Table 2. The experimental results of 3 experiments on vacuum dichroism for pseudoscalar-photon interaction.

| Experiment | Magnetic field $B$ (T) | Magnetic region length $L$ (m) | Effective number $N$ of reflections | Effective $NB^2L^2$ ($T^2 m^2$) | Measured dichroism (nrad) | Derived vacuum dichroism ($10^{-14}$ rad $T^{-2}$ $m^{-2}$) | Pseudo-scalar mass $m_\varphi$ (meV) | Coupling energy scale M ($10^6$ GeV) |
|---|---|---|---|---|---|---|---|---|
| BFRT ($\lambda$: 514 nm) | 2.63-3.87 | 8.8 | 254 | 78,700 | 0.35±0.30 | 0.45±0.38 | for $m_\varphi$ < 0.8 | >2.8 |
| PVLAS ($\lambda$: 1,064 nm) | 5 | 1 | 44,000 | 1,100,000 | 172±22 | 15.6±2.0 | 1-1.5 | 0.2-0.6 |
| Q & A ($\lambda$: 1,064 nm) | 2.3 | 0.6 | 18,700 | 29,900 | -0.4±5.3 | -1±18 | for $m_\varphi$ < 1.7 | > 0.6 |

All 3 experiments in Table 1 aim at detection of vacuum birefringence also. Birefringence of vacuum is predicted by QED (Quantum Electrodynamics) in external transverse electrical or magnetic field (**E** or **B**). For gaseous matter, the birefringence phenomenon is known as Cotton-Mouton effect. The minute Cotton-Mouton effect of gaseous matter in low pressure serves as calibration for these experiments.

Some extensions of the standard model suggest the existence of particles (with mass $m_\varepsilon$) of small unquantized charge $Q_\varepsilon = \varepsilon e$ with $\varepsilon \ll 1$.[25] Recently, Gies, Jaeckel and Ringwald discussed the method of using polarized light propagating in a magnetic field to probe millicharged fermions.[26] Photons with $\omega > 2 m_\varepsilon$ will interact with magnetic field to produce millicharged fermion-antifermion pair. The production cross-section depends on the angle $\theta$ between the transverse magnetic field $B$ and the polarization vector, and the magnetic vacuum is dichroic due to anisotropic absorption. The polarization vector of such an initially linearly polarized photon will rotate by $\Delta\theta$; for small rotation angle,

$$\Delta\theta \approx (1/4) (\kappa_\parallel - \kappa_\perp) L \sin(2\theta), \qquad (6)$$

where $\kappa_\parallel$ and $\kappa_\perp$ are the absorption coefficients corresponding to photon polarizations parallel and perpendicular to $B$ respectively. Let

$$\chi \equiv (3/2) (\omega/m_\varepsilon) (\varepsilon e B/m_\varepsilon^2) = 88.6\ \varepsilon\ (\omega/m_\varepsilon)\ (eV/m_\varepsilon)^2\ (B/T). \qquad (7)$$

Consider the small mass case. For $m_\varepsilon < 0.03$ eV, $\chi \gg 1$, the polarization rotation angle is

$$\Delta\theta \approx (1/4) (\kappa_\parallel - \kappa_\perp) \approx 2.1 \times 10^4\ \varepsilon^{8/3}\ (\omega/\omega_0)^{1/3}\ B^{2/3} L_B, \qquad (8)$$

where $\omega_0$ is the circular frequency corresponding to $\lambda = 1.064$ μm. In Table 3, we compile the experimental results of three experiments on the vacuum dichroism – BFRT, PVLAS and Q & A for millicharged fermion theory with $m_\varepsilon < 0.03$ eV.

Table 3. The comparison of experimental results of 3 experiments on vacuum dichroism with millicharged fermion theory for $m_\varepsilon < 0.03$ eV.

| Experiment | Magnetic field $B$ (T) | Magnetic region length $L$ (m) | Effective number $N$ of reflections | Effective $N B^{2/3} L (\omega/\omega_0)^{1/3}$ ($T^{2/3}$m) | Measured dichroism (nrad) | Derived vacuum dichroism (rad $T^{-2/3}$ $m^{-1}$) | Millifermion charge ratio $\varepsilon$ ($10^{-7}$) |
|---|---|---|---|---|---|---|---|
| BFRT ($\lambda$: 514 nm) | 2.63-3.87 | 8.8 | 254 | 2,650 | 0.35±0.30 | (1.3±1.1)×$10^{-13}$ | 3.5(+1.0, -2.3) |
| PVLAS ($\lambda$: 1,064 nm) | 5 | 1 | 44,000 | 123,500 | 172±22 | (13.9±1.8)×$10^{-13}$ | 8.5±0.4 |
| Q & A ($\lambda$: 1,064 nm) | 2.3 | 0.6 | 18,700 | 18,600 | -0.4±5.3 | (-0.2±2.8)×$10^{-13}$ | 0(+4.6, -0) |

*Principle of experimental measurement* --- The basic principle of experimental measurement is shown as Fig. 1. The laser light goes through a polarizer and becomes polarized. This polarized light goes through a region of magnetic field. Its polarization status is subsequently analyzed by the analyzer-detector subsystem to extract the polarization effect imprinted in the region of the magnetic field. Since the polarization effect of vacuum birefringence and vacuum dichroism in the magnetic field that can be produced on earth is extremely small, one has to multiply the optical pass through the magnetic field by using reflections or Fabry-Perot cavities. BRFT experiment used multiple reflections; PVLAS, Q & A, BMV experiment use Fabry-Perot cavities. For polarization experiment, Fabry-Perot cavity has the advantage of normal incidence of laser light which suppressed the part of polarization due to slant angle of reflections. With Fabry-Perot cavity, one needs to control the laser frequency and/or the cavity length so that the cavity is in resonance. With a finesse of 30,000, the resonant width (FWHM) is 17.7 pm; when rms cavity length control is 10 % of this width, the precision would be 2.1 pm. Hence, one needs a feedback mechanism to lock the cavity to the laser or vice versa. For this, a commonly used scheme is Pound-Drever-Hall method. Vibration introduces noises in the Fabry-Perot cavity mirrors and hence, in the light intensity and light polarization transmitted through the Fabry-Perot cavity. Since the analyzer-detector subsystem detects light intensity to deduce the polarization effect, both intensity noise and polarization noise will contribute to the measurement results. Gravitational-wave community has a long-standing R & D on this. We employ their research results.

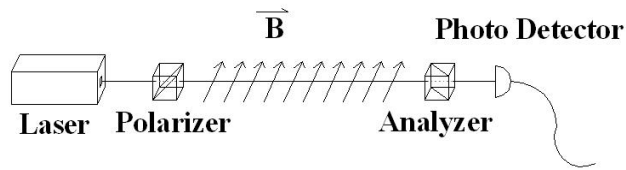

Fig. 1. Principle of vacuum dichroism and birefringence measurement.

The present Q & A experimental schematic is shown in Fig. 2. In section 2, we give some details of the experimental setup. In section 3, we present our polarization detection schemes. In section 4, we address the question of vibration and automatic control. In section 5, we present our measurement procedure and first results. In section 6, we make a couple of discussions and look into the future.

## 2. Experimental set-up of the 3.5 m prototype interferometer

In 1991, we were motivated to do an experiment to measure vacuum birefringence and vacuum dichroism.[27] In 1994, we proposed the Q & A experiment and started to build the experimental facility.[4] In 2002, we finished the first phase of constructing the 3.5 m prototype interferometer and made some Cotton-Mouton coefficient and Verdet coefficient measurements.[28] Starting 2002, we have been in the second phase of Q & A experiment. The schematic of the present setup is shown in Fig. 2. Fig. 3 shows a picture of the experimental apparatus. The 3.5 m prototype interferometer is formed using a high-finesse Fabry–Perot interferometer together with a high-precision ellipsometer. The two high-reflectivity mirrors of the 3.5 m prototype interferometer are suspended separately from two X-pendulum–double pendulum suspensions (basically the design of ICRR, University of Tokyo[29]) mounted on two isolated tables fixed to ground using bellows inside two vacuum chambers. The sub-systems in this second phase are described below.

*2.1. Laser, cavity and finesse measurement*
We use a 1 W diode-pumped 1064 nm CW Nd:YAG laser made by Laser Zentrum Hannover as light source. The laser frequency can be tuned using thermal control and PZT control. We thermal-stabilize the laser at 26.5 °C and use Pound–Drever-Hall technique to lock the laser frequency with the resonant frequency of the 3.5 m Fabry-Perot interferometer. The high-frequency (> 200 Hz) part of the phase error from the photodetector is fed to the PZT to frequency-lock the laser to the cavity resonant frequency; this feedback circuit is a PI$^2$ (proportional-integral$^2$) compensation. The low frequency part (< 200 Hz) of the phase error goes to a DSP based digital control unit to control the cavity length of the interferometer. The laser goes through an isolator, a 12 MHz electro-optic modulator and a mode matching lens to the 3.5 m cavity. We measure the finesse of the 3.5 m cavity by analyzing the ringing profile of transmitted light intensity when the cavity length passing through a resonance. The finesse $F$ is measured to be in the range 28000-32000 in various cases; the uncertainty of measurement is less than 5 %. This finesse corresponds to average number of light passes in the Fabry-Perot cavity to be 17800-20370 ($2F/\pi$). This increases the optical pathlength in the magnetic field of our dichoism/birefringence measurement by 17800-20370 times. A sample data of finesse measurement with fitting to the ringing model[30] is shown in Fig. 4.

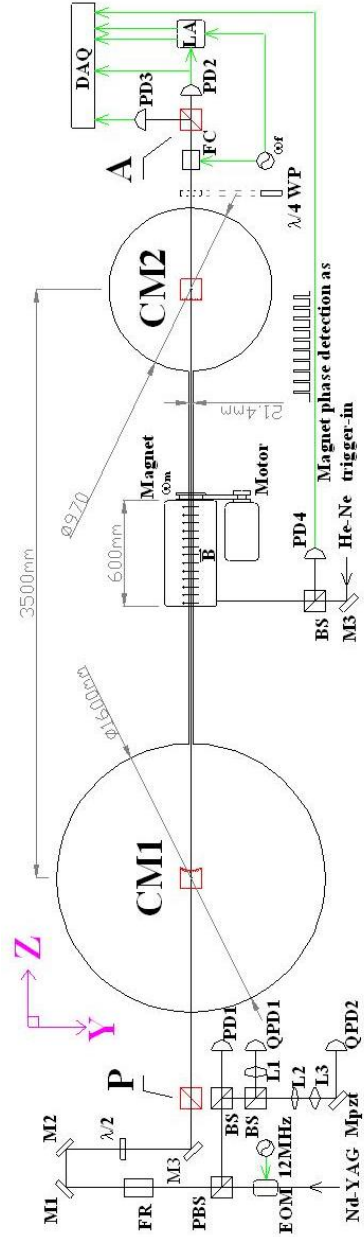

**Figure 2.** The experimental set-up for the present Q & A experiment. Nd-YAG: Nd-YAG laser. EOM: electro-optical modulator. (P)BS: (polarizing) beam splitter. FR: Faraday rotator. M1 ~ M4: mirrors. λ/2: half-wave plate. L1 ~ L3: lens. (Q)PDs: (quadrant) photo-detectors. P: Glan-Taylor type polarizer. A: Glan-Taylor type analyzer. CM1, 2: Fabry-Perot Interferometer (FPI) cavity mirrors. He-Ne: He-Ne laser. λ/4: quarter-wave plate. FC: Faraday cell. LA: lock-in amplifier. DAQ: data acquisition. The polarization rotation signal was modulated by the rotating permanent magnet (up to 10 rev/s) and enhanced by the suspended Fabry-Perot cavity. With an additional modulation by the Faraday cell, the polarization rotation signal can be extracted as a demodulated polarization rotation signal through a lock-in amplifier with precisely detected signal phase for long-term integration when automatic alignment control loop is closed. (Birefringence signal can be detected by inserting a quarter-wave plate before the Faraday Cell to change the birefringence signal to polarization rotation signal.)

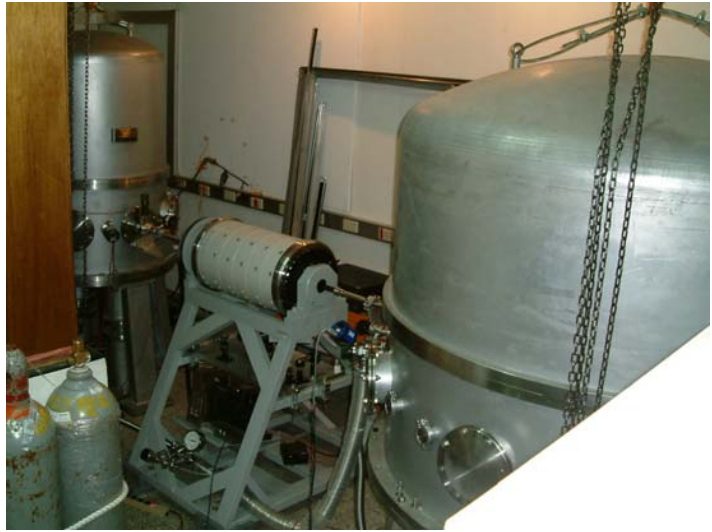

Fig. 3. A picture of experimental apparatus.

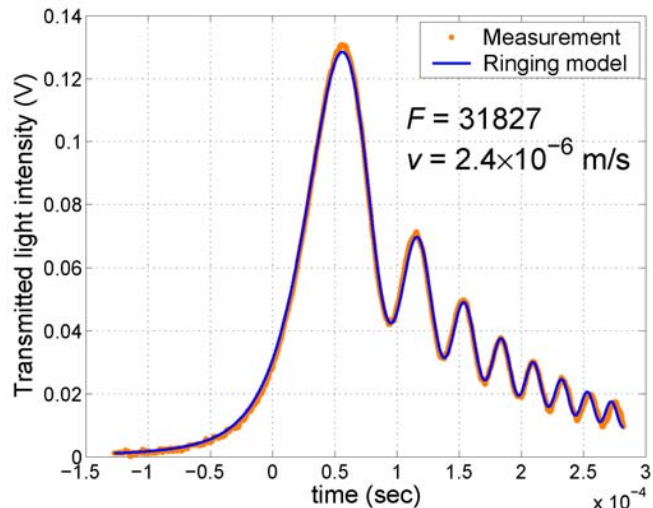

Fig. 4. A finesse measurement with fitting to the ringing model.

*2.2. Mirror Suspension*

Our suspension system for a 3.5 m-cavity mirror consists of an X-pendulum-double-pendulum system (Fig. 5). X-pendulum, named after the crossed-wire structure used in it, was originally designed by Barton, Kuroda, Tatsumi and Uchiyama[29] of the Institute for Cosmic Ray Research, University of Tokyo for TAMA 300 project. We follow closely their design. Our X-pendulum has its resonant frequency around 0.28 Hz[7] and attenuates the horizontal seismic noise in frequency range between 1-20 Hz. The Fabry-Perot cavity mirror is attached to the load table of the X-pendulum via a double pendulum mechanism. The fundamental frequency of the double pendulum is 1.44 Hz.

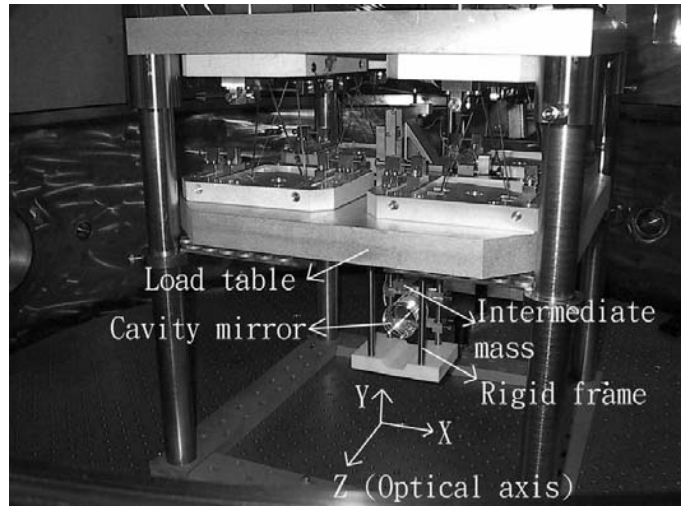

Fig. 5. Picture of one of our X-pendulum-double-pendulum suspension system.

*2.3. Polarizing optics and ellipsometry/polarization rotation measurement*

As shown in Fig. 2, the ellipsometer is formed by the polarizer P, the quarter-wave plate QWP, the polarization modulator Faraday cell FC and the analyzer A. The polarizer P and analyzer A are Glan-Taylor type polarizing prisms which came with other 3 pieces in the same order with a spec of extinction ratio smaller than $10^{-8}$. The extinction ratio of the other 3 polarizing pieces were measured to be $3.8\times10^{-9}$, $1.9\times10^{-9}$, $5.2\times10^{-10}$ with an uncertainty about $10^{-10}$.[31] The measurement data together with fitting for the polarizing piece with extinction ratio $5.2\times10^{-10}$ is shown in Fig. 6.

The Faraday Cell FC is described in reference 28. The modulation response of the Faraday glass was measured to be 0.019 rad/A. The modulation depth for vacuum dichroism measurement is 0.8 mrad (42.1 mA). The polarizer, the analyser and the Faraday cell form a modulated Malus ellipsometer to measure the polarization change. For ellipticity measurement, the quarter-wave plate QWP is inserted before the Faraday Cell FC to change the ellipticity signal to polarization rotation signal.

*2.4. Magnet*

We use a permanent dipole magnet with a 27 mm clear borehole to induce the intracavity dichroism and birefringence of the 3.5 m prototype Fabry-Perot interferometer. The field strength of this magnet at the central axis is 2.3 T with an effective magnetic length 0.6

m. The profile of the magnetic field is shown in Fig. 7.[32, 33] The integrated magnetic field along central axis ∫*Bdl* is 1.15 Tm.

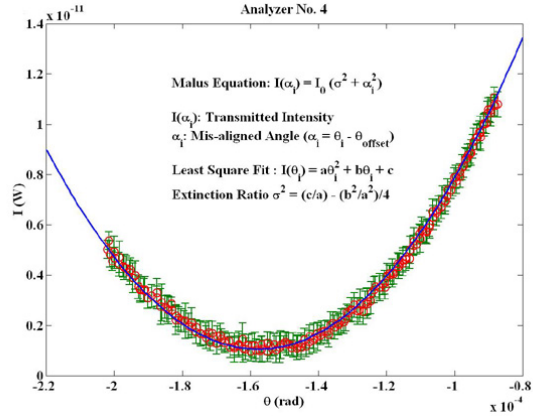

Fig, 6. Data and fitting for the measurement of the extinction ratio of No. 4 analyzer.

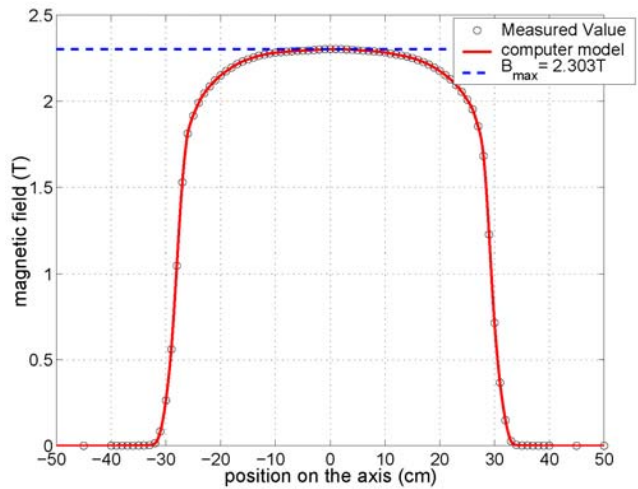

Fig, 7. Measured values of the transverse magnetic field along the central axis of the magnet.

A vacuum tube of ID/OD 21.4 mm/24.6 mm goes through the borehole of the magnet to connect the two mirror-hanging vacuum chambers for laser light to go through.

## 3. Polarization Analyzing Schemes

In this section, we present our polarization detection schemes. As in Fig. 8, the state of polarization of light is modified after light passing through the magnetic field. For example, the Cotton-Mouton effect (CME) of gas transforms a linearly polarized light to an elliptically polarized light[34]; dichroic medium or Faraday glass causes a polarization rotation.[35] In Q & A experiment, the ellipsometer can measure both the ellipticity and rotation signal. Without the quarter-wave plate inserted, the system measures the polarization rotation signal. With the quarter-wave plate inserted, the quarter-wave plate transforms the birefringence signal into polarization rotation signal, and the system measures it. Each optical component in the ellipsometer can be represented by the Jones matrix as listed in Table 4.

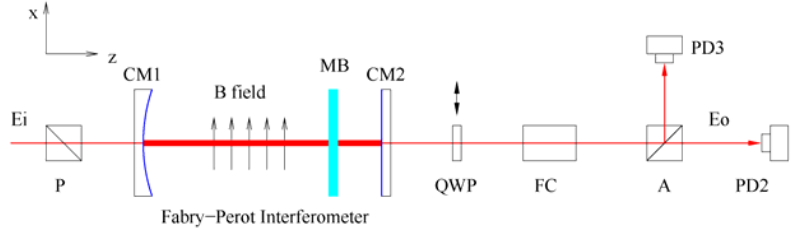

Fig. 8. The schematic of ellipsometer.

Table 4. The Jones matrices for optical components in the ellipsometer

| Optical component | Jones matrix |
|---|---|
| Polarizer P (Transmission axis in the direction of x-axis) | $J_P = \begin{pmatrix} 1 & 0 \\ 0 & 0 \end{pmatrix}$ |
| Analyzer A (Transmission axis in the direction of y-axis) | $J_A = \begin{pmatrix} 0 & 0 \\ 0 & 1 \end{pmatrix}$ |
| Fabry-Perot interferometer with intra-cavity birefringence | $J_{FPI} \cong \dfrac{t_1 t_2}{1-r_1 r_2} \begin{pmatrix} 1+i\dfrac{2F}{\pi}(\psi_0 \cos 2\theta + \varsigma_0 \cos 2\xi) & i\dfrac{2F}{\pi}(\psi_0 \sin 2\theta + \varsigma_0 \sin 2\xi) \\ i\dfrac{2F}{\pi}(\psi_0 \sin 2\theta + \varsigma_0 \sin 2\xi) & 1-i\dfrac{2F}{\pi}(\psi_0 \cos 2\theta + \varsigma_0 \cos 2\xi) \end{pmatrix}$ |
| Quarter-wave plate QWP (Slow axis in the direction of x-axis) | $J_{QWP} = e^{-i\pi/4} \begin{pmatrix} 1 & 0 \\ 0 & -i \end{pmatrix}$ |
| Faraday cell FC | $J_{FC} = \begin{pmatrix} 1 & -\eta \\ \eta & 1 \end{pmatrix}$ |

The Jones matrix $J_{FPI}$ includes the intra-cavity Cotton-Mouton effect (Jones matrix $J_{CME}$) and the residual birefringence of cavity mirrors[36, 37] (Jones matrix $J_{MB}$). The Cotton-Mouton effect can be described by a retardation wave plate whose phase retardation is $2\psi_0$ with slow axis taking an angle $\theta$ with x-axis:

$$J_{CME} = J_R(\theta) \cdot \begin{pmatrix} e^{i\psi_0} & 0 \\ 0 & e^{-i\psi_0} \end{pmatrix} \cdot J_R(-\theta) \\ \cong \begin{pmatrix} 1+i\psi_0 \cos 2\theta & i\psi_0 \sin 2\theta \\ i\psi_0 \sin 2\theta & 1-i\psi_0 \cos 2\theta \end{pmatrix}, \text{ when } \psi_0 \ll 1, \quad (9)$$

; the Jones matrix $J_R(\theta)$ represents polarization (or coordinate) rotation process:

$$J_R(\theta) = \begin{pmatrix} \cos\theta & -\sin\theta \\ \sin\theta & \cos\theta \end{pmatrix}. \tag{9a}$$

Similarly, the residual birefringence of cavity mirrors can also be described by a retardation wave plate MB[36, 37] whose phase retardation is $2\varsigma_0$ with slow axis taking an angle $\xi$ with x-axis:

$$J_{MB} \cong \begin{pmatrix} 1 + i\varsigma_0 \cos 2\xi & i\varsigma_0 \sin 2\xi \\ i\varsigma_0 \sin 2\xi & 1 - i\varsigma_0 \cos 2\xi \end{pmatrix}. \tag{10}$$

The Jones matrix $J_{FPI}$ can be calculated in a similar way as calculating the Airy function of Fabry-Perot interferometer.[35] If the amplitude reflection and transmission coefficients for cavity mirrors are $r_1, t_1$ for CM1 and $r_2, t_2$ for CM2, by using the relation $F = \dfrac{\pi\sqrt{r_1 r_2}}{1 - r_1 r_2} \cong \dfrac{\pi}{2}\dfrac{1 + r_1 r_2}{1 - r_1 r_2}$ (for high finesse mirrors) one can obtain the approximate form of the $J_{FPI}$ as listed in the Table 3.

For ellipticity measurement, the quarter-wave plate is inserted in the light path to change the ellipticity signal to polarization rotation signal in order to beat with the polarization rotation modulation $[\eta = \eta_0 \sin(\omega_f t + \phi_f)]$ provided by the FC. For $E_i = \begin{pmatrix} 1 \\ 0 \end{pmatrix}$, the electric field of light transmitted through the analyzer A is $E_o = J_A \cdot J_{MA} \cdot J_{FC} \cdot J_{QWP} \cdot J_{FPI} \cdot J_P \cdot E_i$, where $J_{MA} = \begin{pmatrix} 1 & -\alpha \\ \alpha & 1 \end{pmatrix}$ represents the analyzer A which is misaligned by a small angle $\alpha$. For simplicity, the misalignment angle of the QWP is not taken into account. By carrying out the Jones-matrix calculation for $E_o$, the light intensity transmitted through the analyzer A is

$$\begin{aligned}
I_o &= E_o \cdot E_o^* \\
&\cong \alpha^2 + \left(\frac{2F}{\pi}\varsigma_0 \sin 2\xi\right)^2 + \frac{1}{2}\eta_0^2 + \frac{1}{2}\left(\frac{2F}{\pi}\psi_0\right)^2 + 2\alpha\left(\frac{2F}{\pi}\varsigma_0\right)\sin 2\xi \\
&\quad + 2\left(\alpha + \frac{2F}{\pi}\varsigma_0 \sin 2\xi\right)\left(\frac{2F}{\pi}\psi_0\right)\sin 2\theta \\
&\quad - \frac{1}{2}\left(\frac{2F}{\pi}\psi_0\right)^2 \cos 4\theta + 2\left(\alpha + \frac{2F}{\pi}\varsigma_0 \sin 2\xi\right)\eta_0 \sin(\omega_f t + \phi_f) \\
&\quad + 2\left(\frac{2F}{\pi}\psi_0\right)\eta_0 \sin 2\theta \sin(\omega_f t + \phi_f) - \frac{\eta_0^2}{2}\cos(2\omega_f t + 2\phi_f).
\end{aligned} \tag{11}$$

Without the QWP, the calculation leads to the light intensity $I_o$:

$$I_o = E_o \cdot E_o^*$$

$$\cong \alpha^2 + \left(\frac{2F}{\pi}\varsigma_0 \sin 2\xi\right)^2 + \frac{1}{2}\eta_0^2 + \frac{1}{2}\left(\frac{2F}{\pi}\psi_0\right)^2$$

$$+ 2\left(\frac{2F}{\pi}\varsigma_0 \sin 2\xi\right)\left(\frac{2F}{\pi}\psi_0\right)\sin 2\theta$$

$$- \frac{1}{2}\left(\frac{2F}{\pi}\psi_0\right)^2 \cos 4\theta \qquad (12)$$

$$+ 2\alpha\eta_0 \sin(\omega_f t + \phi_f)$$

$$+ 2\left(\frac{2F}{\pi}\varsigma_0\right)\left(\frac{2F}{\pi}\psi_0\right)\eta_0 \sin(2\theta + 2\xi)\sin(\omega_f t + \phi_f)$$

$$- \frac{\eta_0^2}{2}\cos(2\omega_f t + 2\phi_f).$$

For magnetic field rotated at a constant speed ($\theta = \omega_m t$), the main Fourier components of the light intensity transmitted through the analyzer A with the QWP inserted and removed are listed in table 5.

Table 5. Main Fourier components of detected signal

| Angular frequency | Amplitude (with the QWP) | Amplitude (without the QWP) |
|---|---|---|
| DC | $\sigma^2 + \alpha^2 + \frac{\eta_0}{2} + \frac{1}{2}\left(\frac{2F}{\pi}\psi_0\right)^2 + \left(\frac{2F}{\pi}\varsigma_0 \sin 2\xi\right)^2 + 2\alpha\left(\frac{2F}{\pi}\right)\varsigma_0 \sin 2\xi$ | $\sigma^2 + \alpha^2 + \frac{\eta_0}{2} + \frac{1}{2}\left(\frac{2F}{\pi}\psi_0\right)^2 + \left(\frac{2F}{\pi}\varsigma_0 \sin 2\xi\right)^2$ |
| $2\omega_m$ | $2\left(\alpha + \frac{2F}{\pi}\varsigma_0 \sin 2\xi\right)\left(\frac{2F}{\pi}\psi_0\right)$ | $2\left(\frac{2F}{\pi}\varsigma_0 \sin 2\xi\right)\left(\frac{2F}{\pi}\psi_0\right)$ |
| $4\omega_m$ | $\frac{1}{2}\left(\frac{2F}{\pi}\psi_0\right)^2$ | $\frac{1}{2}\left(\frac{2F}{\pi}\psi_0\right)^2$ |
| $\omega_f$ | $2\left(\alpha + \frac{2F}{\pi}\varsigma_0 \sin 2\xi\right)\left(\frac{2F}{\pi}\psi_0\right)$ | $2\alpha\eta_0$ |
| $\omega_f \pm 2\omega_m$ | $\left(\frac{2F}{\pi}\psi_0\right)\eta_0$ | $\left(\frac{2F}{\pi}\varsigma_0\right)\left(\frac{2F}{\pi}\psi_0\right)\eta_0$ |
| $2\omega_f$ | $\frac{\eta_0^2}{2}$ | $\frac{\eta_0^2}{2}$ |

From these Fourier amplitudes, the intra-cavity birefringence can be determined. Fourier amplitudes obtained with the QWP inserted is used to determine the ellipticity induced by the Cotton-Mouton effect: $\left(\frac{2F}{\pi}\psi_0\right) = \frac{I_{\omega_f \pm 2\omega_m}}{2I_{2\omega_f}}\eta_0$. The amplitude $I_{\omega_f \pm 2\omega_m}$ without the quarter wave plate inserted is a factor of $\left(\frac{2F}{\pi}\varsigma_0\right)$ smaller than $I_{\omega_f \pm 2\omega_m}$ obtained with the QWP inserted. In BFRT experiment, this factor is 1/300.[2, 38] In our ellipsometer, this factor is found to be 1/670 and, for $F$ = 30,000, corresponds to a residual mirror birefringence of $2\varsigma_0 = 1.6 \times 10^{-7}$ rad (per reflection).

For the ellipsometer with the QWP removed, the ellipsometer is sensitive to the intra-cavity polarization rotation effect (Faraday effect or dichroism).[2] The absorption ratio of dichroic medium for light is dependent on the state of polarization of light. The Jones matrix for dichroic medium whose absorption coefficients are different by $2\beta_0$ in two orthogonal directions can be written as[35]

$$J_{DIC} = J_R(\theta) \cdot \begin{pmatrix} \exp(\beta_0) & 0 \\ 0 & \exp(-\beta_0) \end{pmatrix} \cdot J_R(-\theta)$$
$$\cong \begin{pmatrix} 1 + \beta_0 \cos 2\theta & \beta_0 \sin 2\theta \\ \beta_0 \sin 2\theta & 1 - \beta_0 \cos 2\theta \end{pmatrix} \quad \beta_0 \ll 1. \tag{13}$$

By the same calculation procedure, similar Fourier amplitudes of light intensity transmitted through the analyzer with the QWP removed can be obtained:

$$I_{\omega_f \pm 2\omega_m} = \left(\frac{2F}{\pi}\beta_0\right)\eta_0; \quad I_{2\omega_f} = \frac{\eta_0^2}{2}. \tag{14}$$

Therefore, the quantity $\beta_0$ can be determined as follows:

$$\beta_0 = \frac{\pi}{2F}\frac{I_{\omega_f \pm 2\omega_m}}{2I_{2\omega_f}}\eta_0. \tag{15}$$

### 4. Vibration Isolation and Automatic Length Control

Each mirror of the Fabry-Perot cavity is suspended by an X-pendulum with a double pendulum as the second stage, as shown in Fig. 5. The isolation ratio of the X-pendulum is about 20 dB at 1Hz, 50 dB at 4 Hz, 28 dB at 8 Hz and 34 dB at 20 Hz. The double pendulum consists of an intermediate mass, a recoil mass RM and the mirror. The longitudinal length of the Fabry-Perot cavity can be adjusted by applying force to magnets on the mirror through coils held by the recoil mass RM. A 0.6 m long rotating permanent dipole magnet with a maximum central field of 2.3 T is located between front mirror and end mirror for producing a polarized vacuum. For a more detailed exposition of vibration isolation and automatic length control, please see reference 7 and 39. During the automatic length control feed back, the cavity length is controlled to a precision of 2.1 pm rms to follow the resonance. The full range of coil control is about 2 μm.

To test our suspension system, we have measured the background seismic noise and the relative displacement control noise between two cavity mirrors using heterodyne interferometry.[40] For seismic noise measurement, we simply measured the displacement of the suspended cavity mirror relative to a reference point on the optical table on which the suspension system is mounted; the measured result is the resonant displacement of the suspension itself relative to the seismic motion on the optical bench. For measuring relative displacement between cavity mirrors, a diode-pumped Nd:YAG laser was frequency stabilized to the suspended Fabry-Perot cavity via the Pound-Drever-Hall locking technique. The relative displacement ($>$ 1 Hz) is estimated from the correction signal that drives the PZT of the laser. The spectral density of these displacements are plotted in figure 3 of reference 7, it shows that our suspension system is capable of isolating seismic noise in frequency range between 1-20 Hz. There are many small resonant modes in this frequency band; however, we can avoid these peaks by choosing suitable modulation frequency of the magnetic field. The noise floor at 5-10 Hz is 2 orders of magnitude below that at 0.05 Hz.

## 5. Measurement Procedure and First Results

The ellipsometer is calibration-checked by performing the measurement of gaseous Cotton-Mouton effect (CME). The ellipticity due to CME is written as

$$\Psi_0 = \frac{2F}{\lambda} \cdot \Delta n(\mathrm{N}_2) \cdot \int B^2(l) dl \cdot \frac{P_{gas}}{1\,\mathrm{atm}}, \qquad (16)$$

where $P_{gas}$ is the pressure of nitrogen gas, $\lambda$ is the wavelength of light, and $\Delta n(\mathrm{N}_2) = n_{\parallel} - n_{\perp}$ is the difference between index of refraction parallel and perpendicular to the magnetic field per unit pressure per unit magnetic field squared. The measurement of nitrogen CME is taken at several different pressures (room temperature ~ 19.5º C). By defining $\Psi_N$ and re-writing (16) as

$$\Psi_N = \frac{\Psi_0 \cdot (1\,\mathrm{atm})}{\frac{2F}{\lambda} \int B^2(l) dl} = \Delta n(\mathrm{N}_2) \cdot P_{gas}, \qquad (17)$$

$\Delta n(\mathrm{N}_2)$ can be determined by fitting the data to a straight line, as shown in figure 9. The measured birefringence in nitrogen is $\Delta n(\mathrm{N}_2) = -(2.66 \pm 0.12) \times 10^{-13}$ T$^{-2}$ atm$^{-1}$. The Cotton-Mouton Coefficient (defined at pressure of 1 atm) is $C_{CM} = \Delta n(\mathrm{N}_2)/\lambda = -(2.50 \pm 0.11) \times 10^{-7}$ T$^{-2}$ m$^{-1}$. The measurement was done with low-finesse cavity mirrors (finesse ~4400). This result is consistent with the measured values in the literature.[34]

    We have also made a few nitrogen CME measurements with high-finesse cavity mirrors. Fig. 11 is the spectrum of the measured ellipticity and figure 11 is the polar plot of the amplitude and phase at twice the magnet rotation frequency ($2\omega_m$). In the polar plot, the phase depends on the alignment of the QWP, directions of magnet rotation and FC polarization rotation, and the phase of -180 deg means that the fast axis of nitrogen's CME is parallel to the magnetic field. In this run the finesse is 25,700, and according to (17) the Cotton-Mouton Coefficient in this run is $C_{CM} = -2.1 \times 10^{-7}$ T$^{-2}$ m$^{-1}$ which is roughly consistent with the result obtained with the low finesse cavity mirrors.

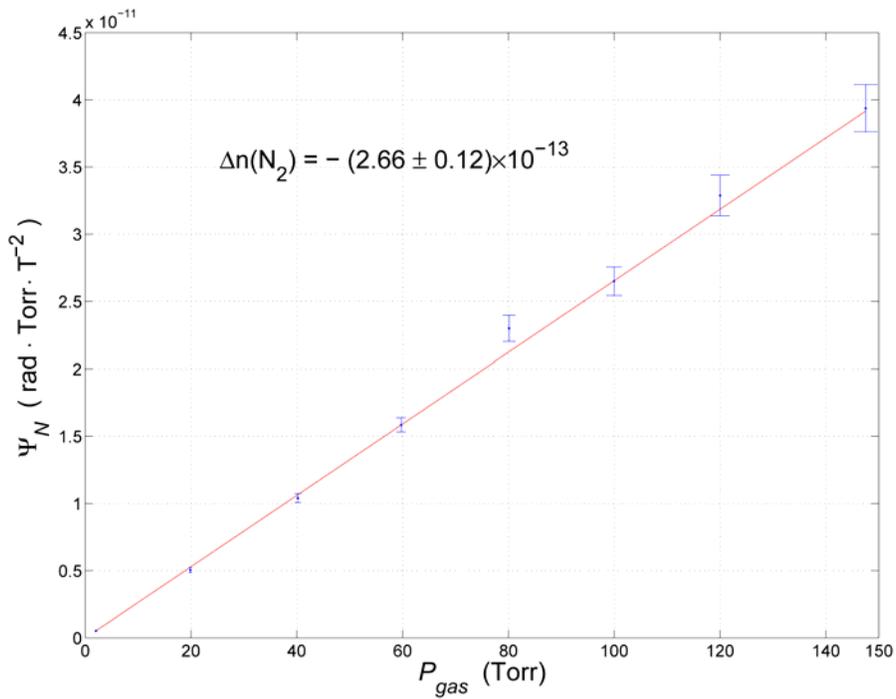

Fig. 9. The Cotton-Mouton effect measurement of nitrogen gas.

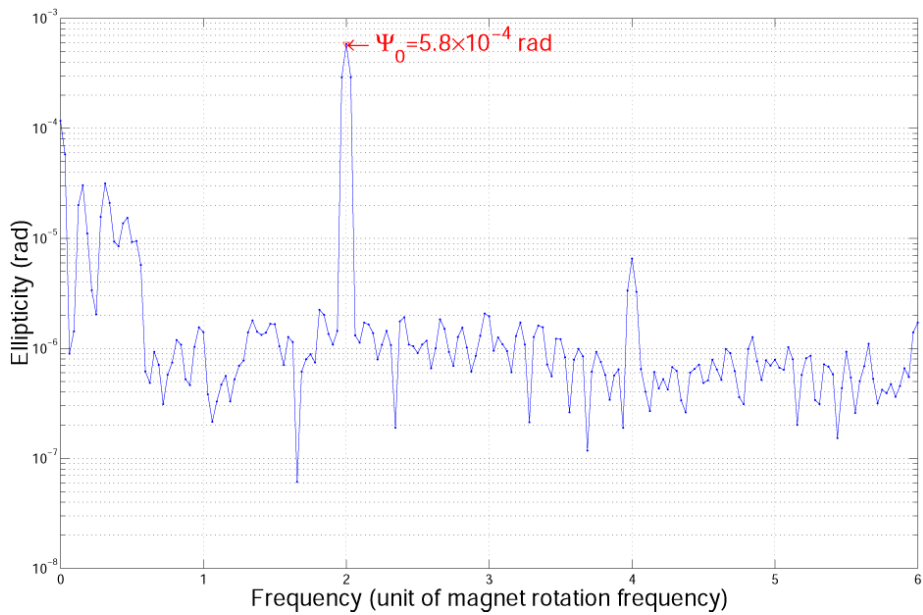

Fig. 10. CME of 15 Torr nitrogen gas: the spectrum of measured ellipticity

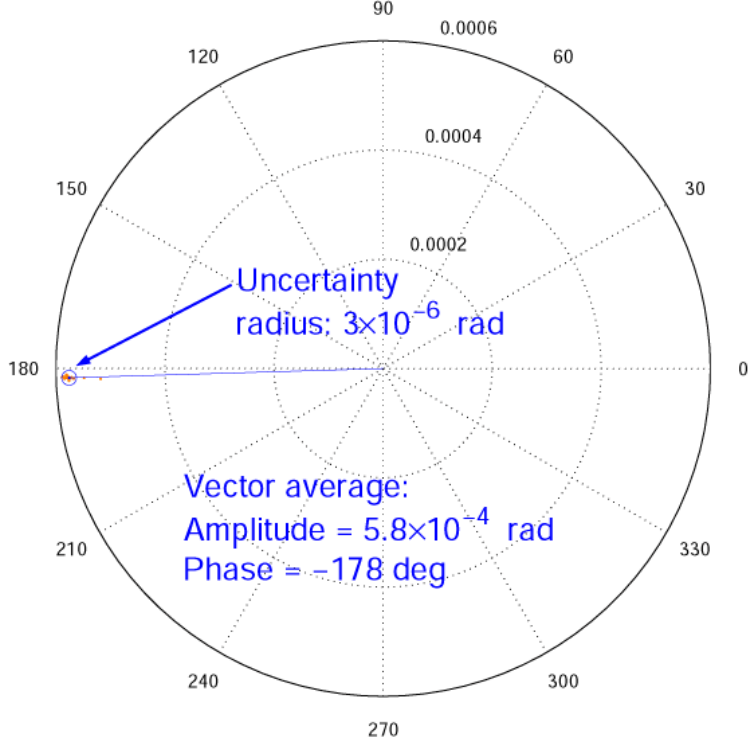

Fig. 11. CME of 15 Torr nitrogen gas: amplitude and phase at twice the magnet rotation frequency ($2\omega_m$)

To measure the vacuum polarization rotation effect, the QWP is removed. Light transmitted through the analyzer A is received by the photo-detector PD2. The output voltage of the PD2 is sent into a lock-in amplifier (Stanford Research System, SR830) for detecting the Fourier component at $\omega_f$. According to equation (12) in section 3, this Fourier component has the following form:

$$I_{demo} = 2\left[\alpha + \left(\frac{2F}{\pi}\varsigma_0\right)\left(\frac{2F}{\pi}\psi_0\right)\sin(2\omega_m t + 2\xi) + \left(\frac{2F}{\pi}\right)\beta_0 \sin(2\omega_m t + \phi_\beta)\right]\eta_0$$

(18)

(the suffix *demo* stands for demodulation), where $\phi_\beta$ is introduced to be general. To make the dichroic effect $\beta_0$ the dominant term in eq. (18), the term $\left(\frac{2F}{\pi}\varsigma_0\right)\left(\frac{2F}{\pi}\psi_0\right)$ has to be suppressed by pumping out the gas from the interaction region. Let us estimate the polarization rotation effect observed by the PVLAS collaboration[1] in the Q & A experiment. Since the effect is scaled by $B^2 L^2$, we have

$$\frac{1.32\,(\text{T}^2\text{m}^2)}{25\,(\text{T}^2\text{m}^2)} \cdot (3.9 \pm 0.5) \times 10^{-12} \cdot \frac{2F}{\pi} = (3.9 \pm 0.5) \times 10^{-9} \quad (\text{rad}).\qquad(19)$$

For nitrogen at 1 mTorr, $F = 30{,}000$ and $\left(\dfrac{2F}{\pi}\varsigma_0\right) = \dfrac{1}{670}$, the term $\left(\dfrac{2F}{\pi}\varsigma_0\right)\left(\dfrac{2F}{\pi}\psi_0\right)$ we are estimating is

$$\left(\frac{2F}{\pi}\varsigma_0\right)\left(\frac{2F}{\pi}\right)\pi \cdot C_{CM} \cdot \int B^2(l)dl \cdot \frac{1 \times 10^{-3}}{760}\frac{1}{670} = 8 \times 10^{-11}\ \text{rad}\qquad(20)$$

which is about 2% of the polarization rotation effect. The gas pressure during this experimental run was kept lower than 1 mTorr.

The angle $\theta$ is measured by 32 equally spaced reflection mirrors glued on a ring fixed to the magnet. A 633 nm He-Ne laser is steered to shine the ring, and the 633 nm laser beam reflected by the 32 mirrors is received by a photo-receiver. When the magnet is rotated, the output of the photo-receiver is an analog pulse train, where each pulse corresponds to a rotation angle of 11.25 deg. One of the 32 reflection mirrors is partially covered by a piece of opaque tape which makes the width of corresponding analog pulse the narrowest. The rising edge of this narrowest analog pulse occurs when the magnetic field B is in angle 95.1 deg with the polarization vector of input laser (1064 nm).

Signals including the outputs of the PD2, PD3, lock-in amplifier, and the analog pulse train were sampled at 20 kHz and saved as files for off-line analysis. The data files were first processed by a C language routine. The routine first checks the lock status of the Fabry-Perot interferometer. If the FPI is locked, it begins to re-sample the output of lock-in amplifier, i.e. the signal $I_{demo}$, according to the rising edge of the analog pulse train. This re-sampling process always begins at the rising edge of the narrowest analog pulse in order to maintain the phase relation between each data segment. By using the fourier component $I_{2\omega_f}$, the re-sampled signal is converted to unit of radian and written as

$$I_{rs} = \alpha(\theta_i) + \left(\frac{2F}{\pi}\varsigma_0\right)\left(\frac{2F}{\pi}\psi_0\right)\sin(2\theta_i + 2\xi) + \left(\frac{2F}{\pi}\right)\varepsilon_0 \sin(2\theta_i + \phi_\beta),\qquad(21)$$

where

$$\theta_i = 95.1 + 11.25 \cdot i \quad (\text{deg.}) \qquad\qquad i = 0,1\cdots,31.\qquad(22)$$

The spectrum of the $I_{rs}$ in this run is shown in figure 12; the integration time for this spectrum is 18.9 hr. Total data acquisition time is 19.5 hr, therefore the achieved duty cycle is 96.9%. No obvious peak is seen at $2\omega_m$ in figure 12. To detect the amplitude and phase of the Fourier component at $2\omega_m$ in $I_{rs}$, $I_{rs}$ has to be demodulated (lock-in detection) again by the magnet rotation angle $\theta = \omega_m t$. In this demodulation, a set of orthogonal signal is formed by $P = I_{rs} \cdot \sin 2\omega_m t$ and $Q = I_{rs} \cdot \cos 2\omega_m t$. The DC terms of $P$ and $Q$ are

$$P_{DC} = \left(\frac{2F}{\pi}\right)\beta_0 \cos\phi_\beta \qquad (23)$$

and

$$Q_{DC} = \left(\frac{2F}{\pi}\right)\beta_0 \sin\phi_\beta . \qquad (24)$$

The amplitude and phase at $2\omega_m$ can be determined from (23) and (24):

$$\left(\frac{2F}{\pi}\right)\beta_0 = \sqrt{P_{DC}^2 + Q_{DC}^2} , \qquad (25)$$

$$\phi_\beta = \arctan\frac{Q_{DC}}{P_{DC}}. \qquad (26)$$

The obtained amplitude and phase at $2\omega_m$ is plotted in the polar axis as shown in figure 13. By vector average, the averaged vector (color blue) has amplitude of 5.2 nrad and phase of -94.3 deg. For dichroism induced rotation, the phase is 0 deg according to (13). The in phase dichroism quadrature amplitude is -0.4 nrad. The average of the Fourier amplitudes near 2 $\omega_m$ is 5.3 nrad. This is the noise floor of our measurement. Hence, we have (-0.4 ± 5.3) nrad for the measured dichroism. The systematic error is less then 20 % and will not be discussed here.

For $\left(\frac{2F}{\pi}\right)\beta_0 = 5.3$ nrad, by (3) we can give a limit on the parameter *M*, the axion-photon interaction energy scale:

$$M > \left(\frac{NB^2L^2}{16\beta_0}\right)^{1/2} = 6.0\times 10^5 \text{ GeV}, \qquad (27)$$

valid for

$$m_\phi < \left(\frac{2\pi\omega}{L}\right)^{1/2} = 1.7\,\text{meV}. \qquad (28)$$

This constraint is listed in Table 2.

For the constraint on induced vacuum dichroism in millicharged fermion theory, please see Table 3 and discussions in Section I.

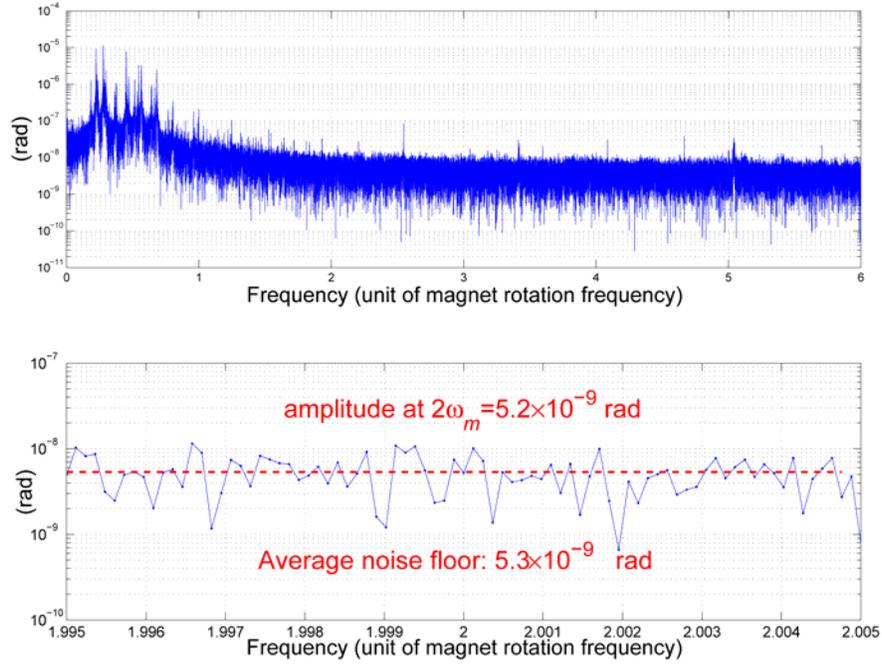

Fig. 12. The spectrum of the re-sampled signal $I_{rs}$

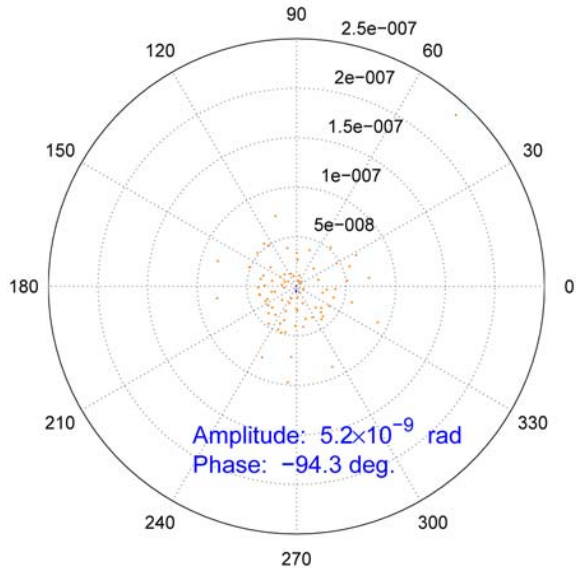

Fig. 13. The polar plot of Fourier component at $2\omega_m$ of the signal $I_{rs}$.

## 6. Discussion and Outlook

We have presented our first results of the Q & A experiment. At the present accuracy, there is no indication of pseudoscalar-photon interaction and existence of millicharged fermions. For pseudoscalar-photon interaction, our accuracy needs to be improved in order to confirm/check the PVLAS results. For the existence of millicharged fermions, our first results have similar accuracy compared to PVLAS and BFRT experiment; for the small fermion mass $m_\varepsilon < 0.03$ eV region, our experiment gives an upper limit $4.6 \times 10^{-7}$ on the millicharged fermion charge ratio $\varepsilon$.

We intend to extend our interferometer to 7 m in arm length and use a 5 m rotating permanent magnet with transverse field 2.5 T in our experiment. This way we will gain a factor of 10 in accuracy in search of (pseudo)scalar-photon interaction and a factor of 100 in accuracy in search of millicharged fermions. We intend also to improve our ellipticity detection sensitivity and use frequency-doubled Nd:YAG in order to reach QED vacuum birefringence sensitivity.

Astrophysical bounds in the literature are more stringent. There are recent studies on whether these bounds could be evaded.[41-44] Readers are referred to the cited references for the current situation.

Michelson-Morley experiment[45] on the comparison of light velocity in different directions is crucial for the development of modern physics. Experiments on light propagation for different polarizations may or may not confirm vacuum dichroism; in either situation, it is important to find out.

## Acknowledgements

We thank the National Science Council (NSC 93-2112-M-007-022, NSC 94-2112-M-007-012, NSC 95-2112-M-007-036) for supporting this program. We would also like to thank J.-T. Shy for lending equipments and J.-P. Wang for encouragements during the long period of building this experiment.

## References

1. E. Zavattini et al., *Phys. Rev. Lett.* **96**, 110406 (2006).
2. R. Cameron, *et al. Phys. Rev.* **D47,** 3707 (1993).
3. R. Pengo, *et al*, in *Frontier Test of QED and Physics of the Vacuum*, ed. E. Zavattini, *et al.* (Sofia: Heron Press, 1998) p. 59; and references there in.
4. W.-T. Ni, *Frontier Test of QED and Physics of the Vacuum*, ed. E. Zavattini, *et al.* (Sofia: Heron Press, 1998) p. 83; W.-T. Ni, *Chin. J. Phys.* **34,** 962 (1996); and references there in.
5. F. Nezrick, *Frontier Test of QED and Physics of the Vacuum*, ed. E. Zavattini, *et al.* (Sofia: Heron Press, 1998) p. 71; and references there in.
6. Askenazy S, Billette J, Dupre P, Ganau P, Mackowski J, Marquez J, Pinard L, Portugall O, Ricard D, Rikken G L J A, Rizzo C, Trenec G and Vigue J, *Quantum Electrodynamics and Phisics of the Vacuum*, ed. G Cantatore (AIP, 2001) p. 115.
7. S.-J. Chen, S.-H. Mei and W.-T. Ni, *Improving ellipticity detection sensitivity for the Q & A vacuum birefringence experiment*, hep-ex/0308071 (2003).
8. W.-T. Ni, A Nonmetric Theory of Gravity, preprint, Montana State University, Bozeman, Montana, USA (1973). The paper is available via http://gravity5.phys.nthu.edu.tw/webpage/article4/index.html.
9. W.-T. Ni, *Bull. Am. Phys. Soc* **19**, 655 (1974).
10. W.-T. Ni, *Phys. Rev. Lett.* **38**, 301 (1977).
11. S. M. Carroll, G. B. Field, R. Jackiw, *Phys. Rev.* **D 41**, 1231 (1990).
12. S. M. Carroll and G. B. Field, *Phys. Rev.* **D 43**, 3789 (1991).
13. W.-T. Ni, *Chin. Phys. Lett.* **22**(1), 33-35 (2005).
14. W.-T. Ni, *Int. J. Mod. Phys.* **D 13,** 901 (2005); this reference gives a brief account of previous works on the cosmological polarization-rotation effect.
15. B. Feng, M. Li, J.-Q. Xia, X, Chen, X. Zhang, *Phys.Rev.Lett.* **96**, 221302 (2006); this paper gives a value of $\Delta\varphi = 6 \pm 6$ deg = $0.1 \pm 0.1$ (rad) which is consistent with a limit of 0.1 of $\Delta\varphi$ of reference 13.
16. S. Weinberg, Phys. Rev. Lett. **40**, 233 (1978).

17. F. Wilczek, Phys. Rev. Lett. **40**, 279 (1978).
18. R. D. Peccei and H. R. Quinn, Phys. Rev. Lett. **38**, 1440 (1977).
19. J. Kim, Phys. Rev. Lett. **43**, 103 (1979).
20. M. Dine *et al.,* Phys. Lett. **104B**, 1999 (1981).
21. M. Shifman *et al.,* Nucl. Phys.**B166**, 493 (1980).
22. P. Sikivie, *Phys. Rev. Lett.* **51,** 1415 (1983); A. A. Anselm, *Yad. Fiz.* **42,** 1480 (1985); M. Gasperini, *Phys. Rev. Lett.* **59,** 396 (1987).
23. L. Maiani, R. Petronzio and E. Zavattini, *Phys. Lett.* **B175,** 359 (1986).
24. G. Raffelt and L. Stodolsky, *Phys. Rev.* **D37,** 1237 (1988).
25. B. Batell and T. Gherghetta, *Phys. Rev.* **D73,** 045016 (2006); and reference there in.
26. H. Gies, J. Jaeckel, and A. Ringwald, Polarized light propagating in a magnetic field as a probe of millicharged fermions, DESY 06-105, arXiv:hep-ph/0607118 v 1 (2006).
27. W.-T. Ni, *et al., Mod. Phys. Lett.* **A6,** 3671 (1991).
28. J.-S. Wu, S.-J. Chen and W.-T. Ni, *Class. Quantum Grav.* **21,** S1259 (2004).
29. D. Tatsumi, M. A. Barton, T. Uchiyama and K. Kuroda, *Rev. Sci. Instrum.*, **70,** 1561 (1999); and references therein.
30. J. Poirson, F. Bretenaker, M. Vallet, and A. Le Floch, *J. Opt. Soc. Am. B*, **14**, 2811 (1997): and references there in.
31. H.-H. Mei, S.-J. Chen and W.-T. Ni, *J. of Phys.: Conf. Series*, **32**, 236 (2006).
32. Z. Dong, W.-T. Ni, W. Yang, and Z. Wang, Prototype magnet design for magnetic birefringence of vacuum – Q & A experiment (in Chinese with an English abstract), *Advanced Technology of Electrical Engineering and Energy*, **23**(4), 65 (2004).
33. Z. Wang, W. Yang, T. Song, and L. Xiao, *IEEE Transactions on Applied Superconductivity* **14, (2)** 1264-1266 (2004).
34. C. Rizzo, A. Rizzo and D. M. Bishop, *Int. Rev. Phys. Chem.* **16,** 81; and reference there in (1997).
35. F. Pedrotti and L. Pedrotti, *Introduction to Optics*, 2nd edn. (Prentice-Hall, New York, 1993).
36. D. Jacob *et al Opt. Lett.* **20**, 671-673 (1995).
37. F. Brandi *et al.*, *Appl. Phys.* B **65** 351-355 (1997)
38. R. Cameron, Search for new photon couplings in a magnetic field, Ph.D. thesis (University of Rochester, 1992)
39. S.-J. Chen, H.-H. Mei and W.-T. Ni, *J. of Phys.: Conf. Series*, **32**, 244 (2006).
40. H. C. Yeh, W. T. Ni and S. S. Pan, *Int. J. Mod. Phys.* **D11,** 1087 (2002).
41. E. Masso and J. Redondo, JCAP **0509**, 015 (2005).
42. P. Jain and S. Mandal, astro-ph/0512155.
43. J. Jaeckel, E. Masso, J. Redondo, A. Ringwald and F. Takahashi, hep-ph/0605313.
44. E. Masso and J. Redondo, hep-ph/0606163.
45. A. A. Michelson and E. W. Morley, *Am. J. Sci.* **34**, 333 (1887).